\documentclass[prl,twocolumn,floatfix,superscriptaddress]{revtex4}
\usepackage{amssymb,amsmath}
\usepackage{graphicx}
\usepackage{subfigure}

\newcommand{\mb}[1]{\ensuremath{\mathbf{#1}}}

\begin{document}
\title{Interface Width and Bulk Stability: requirements for the
  simulation of Deeply Quenched Liquid-Gas Systems}


\author{A.J. Wagner} 
\affiliation{Department of Physics, North Dakota State University,
Fargo, ND 58105}
\email{alexander.wagner@ndsu.edu}
\author{C.M. Pooley}
\affiliation{Rudolf Peierls Centre for Theoretical Physics, Oxford University, OX1 3NP, U.K.}

\begin{abstract}
Simulations of liquid-gas systems with extended interfaces are
observed to fail to give accurate results for two reasons: the
interface can get ``stuck'' on the lattice or a density overshoot
develops around the interface. In the first case the bulk densities
can take a range of values, dependent on the initial conditions. In
the second case inaccurate bulk densities are found.  In this
communication we derive the minimum interface width required for the
accurate simulation of liquid gas systems with a diffuse interface. We
demonstrate this criterion for lattice Boltzmann simulations of a van
der Waals gas. When combining this criterion with predictions for the
bulk stability we can predict the parameter range that leads to stable
and accurate simulation results. This allows us to identify parameter
ranges leading to high density ratios of over 1000. This is despite
the fact that lattice Boltzmann simulations of liquid-gas systems were
believed to be restricted to modest density ratios of less than 20
\cite{Inamuro}.
\end{abstract}

\maketitle 

Application of lattice Boltzmann methods to the simulation of
liquid-gas systems has been one of the early successful applications
of lattice Boltzmann. Two very different algorithms were developed to
do this: Swift et. al. \cite{swift,Holdych,inamuro,kalarakis}
developed an algorithm based on implementing a pressure tensor and
Shan et al. \cite{shan,he,luo} developed an algorithm based on
mimicking microscopic interactions. These algorithms have been
succesfully applied to simulations of phase-separation \cite{mecke},
drop-collisions \cite{Inamuro,jia}, wetting dynamics and spreading
\cite{Kusumaatmaja}, and the study of dynamic contact angles
\cite{grubert,briant,Kwok}. Only recently has it been shown that both
algorithms perform very similarly when higher order corrections are
taken into account \cite{pre}.

Previously there have been only heuristic analysis of what range of
paramteres lead to accurate simulation results. In some parameter
ranges not too far from the critical point too thin interefaces can
lead to interfaces that stick to the lattice. This can lead to
non-unique bulk densities for the liquid and gas phases that depend
on the inital contitions. Further from the critical point density
overshooting is observed at the interfaces. This also leads to
incorrect bulk densities. In this communication we present a criterion
that allows us to predict the range of acceptable values for the
interface width. We will show that the accuracy of the algorithm
rapidly deteriorates when this limit is exceeded. The second important
contribution of this communication is a more general definition of the
equation of state. Usual lattice Boltzmann methods recover the ideal
gas equation of state with a pressure of $p=\rho/3$ when the gas is
dilute. While relaxing this requirement has no effect on the
interfacial properties it allows us to adjust the bulk-stability of
the lattice Boltzmann method.  Taken together this allows us to
determine sets of paramters for which very deep quenches can be
simulated. This is an important result since lattice Boltzmann
methods were previously believed to be limited to density ranges of
about 20. We can now identify parameter ranges for which we can obtain
much larger density ratios. In this communication we present some
simulations with density ratios larger than 1000.

The key to a successful simultion of liquid-gas systems is the
faithful representation of the interface. In particular we need to
obtain a constant pressure across a flat interface. For a standard Landau
Free energy of $F=\int \psi_0+\kappa/2\;(\nabla\rho)^2$, where $\rho$ is the density, we
have
\begin{equation}
P_{\alpha\beta}=\left[p_0(\rho) - \kappa (\rho\nabla^2\rho + \frac{1}{2}\nabla\rho.\nabla\rho)\right]\delta_{\alpha\beta}+\kappa \nabla_\alpha\rho \nabla_\beta\rho  
\end{equation}
where $p_0=\rho\partial_\rho\psi_0-\psi_0$ is the bulk pressure.  In equilibrium the normal
pressure is constant across an interface. For the equilibrium profile
$\rho(x)$ corresponding to a bulk pressure of $p_B$ this implies
\begin{equation}
\kappa=\frac{p(\rho)-p_B}{\rho\partial_x^2\rho-\frac{1}{2}\partial_x\rho \partial_x\rho}.
\end{equation}
A problem arises for discrete simulation methods because the density
derivatives have to be replaced by discrete derivatives. For
simplicity we will limit our analysis to the one-dimensional case
here. The standard definitions for the discrete derivative are
$\partial_x\rho=0.5[\rho(x+1)-\rho(x-1)]$ and $\partial_x^2\rho=\rho(x+1)+\rho(x-1)-2\rho(x)$. This puts
severe limits on the allowable values for $\kappa$.

We can now perform a simple estimate of the
minimum value $\kappa_{m}$ that allows this pressure to be the equilibrium
pressure. For any point on the interface with density $\rho_s$ we
consider two neighboring points, one with a smaller density $\rho_-$ and
one with a larger density $\rho_+$. 
If $\rho_l$ is the gas density and $\rho_g$ is the gas
density we deduce a lower limit for the
smallest possible value $\kappa_{m}(\rho_s)$ by varying the values of $\rho_+$
and $\rho_-$, but not allowing overshooting:
\begin{equation}
\kappa_{m}(\rho_s)=\min_{
\begin{array}{c}
\scriptstyle \rho_g<\rho_-<\rho_s\\
\scriptstyle \rho_s<\rho_+<\rho_l
\end{array}}\frac{p(\rho)-p_b}{\rho_s(\rho_--2\rho_s+\rho_+)-\frac{1}{8}(\rho_+ -\rho_-)^2}.
\label{kappamin}
\end{equation}
Since we need to allow any value of $\rho_s$ between $\rho_g$ and $\rho_l$ the
minimum allowable value of $\kappa$ is then given by
\begin{equation}
\kappa_m=\max_{\rho_g<\rho_s<\rho_l}\kappa_m(\rho_s).
\end{equation}

To be concrete and to illustrate the power of the minimum $\kappa$
prediction we will consider the pressure of a van der Waals gas
\begin{eqnarray}
p(\rho)&=&p_0\left(\frac{\rho}{3-\rho}-\frac{9}{8}
  \rho^2\theta_c\right)
\label{VdW_int_new}
\end{eqnarray}
Please note that we introcuded a scale-factor $p_0$ with the
pressure. Such a scale factor clearly is not expected to change the
equilibrium behavior. However, as we will see below, it can have a
profound effect on the bulk stability of the system. Most previous
aproaches correspond to a choice of $p_0=1$ in
Eq. (\ref{VdW_int_new}). Only for this choice will the lattice
Boltzmann method recover the standard lattice Boltzmann method for
ideal gases in the limit of small densities.

A good approximation for the interface shape that becomes exact close
to the critical point is given by
\begin{equation}
\rho^{init}(x)=\rho_g+\frac{\rho_l-\rho_g}{2}\left[1+  
\tanh\left(\frac{x}{w(\kappa,\theta/\theta_c)}\right)\right].
\label{profile}
\end{equation}
where $\rho_l$ and $\rho_g$ are the equilibrium gas and liquid densities. 
The interface width is given by
\begin{equation}
w(\kappa,\theta/\theta_c)=\sqrt{\frac{2\kappa}{\theta_c/\theta-1}}
\label{width}
\end{equation}
This profile is not the exact analytical solution to the differential
equation $\nabla P=0$, but it is very close to it. 

\begin{figure}
\includegraphics[width=0.9\columnwidth]{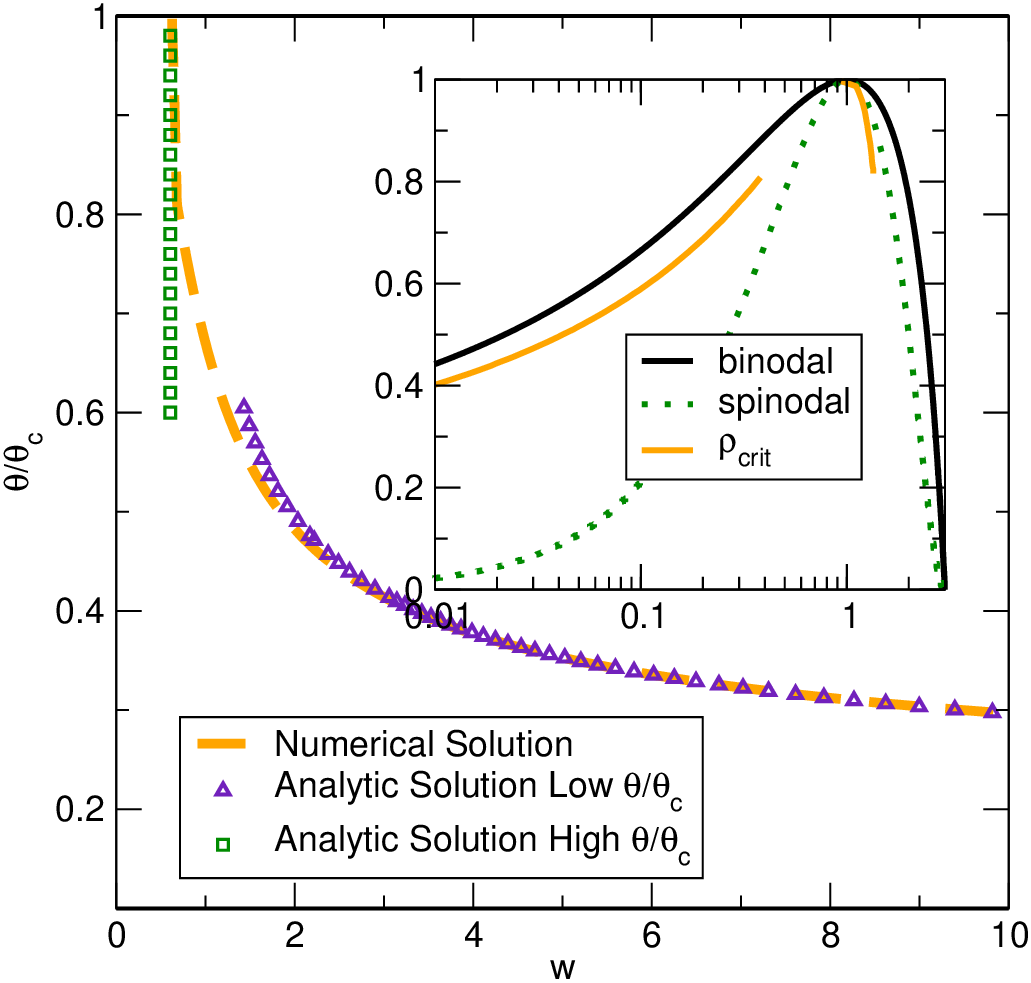}
\caption{(Color online) The two limiting cases for which we can obtain an analytical
  approximation to the $w(\kappa_{m})$ relation. The inset shows the value
  of the critical density $\rho_{crit}$, which is the value $\rho_s$ takes
  when (\ref{kappamin}) is maximized.  Note that there is a
  discontinuity.}
\label{thetaw}

\end{figure}

We first examine numerically which values of $\rho_s$ in Eq. (\ref{kappamin}) lead
to the most restrictive constraint, \textit{i.e.} the largest value of
$\kappa_m(\rho_s)$.  The orange line in the inset of Figure \ref{thetaw} shows
how this density $\rho_{crit}$ varies as a function of temperature.  Near
to the critical temperature, the orange line in the inset in Figure
\ref{thetaw} lies close to the high density spinodal curve.  This is
because $P-p_b$ has its highest magnitude here, therefore helping to
maximize $\kappa_{m}$ within this region.  An analytical estimate for
$\kappa_{m}$ can be obtained by expanding the pressure around the critical
density, giving
\begin{eqnarray} 
P-p_b = -\frac{9}{4} \left( \theta_c-\theta \right) \left(\rho_s-1 \right) + \frac{3}{16} \left(\rho_s-1 \right)^3.
\end{eqnarray}
Within this regime, $P-p_b$ is large and negative, therefore $\rho_-$ and
$\rho_+$ must be chosen to make the denominator in (\ref{kappamin}) as
negative as possible.  A suitable choice is $\rho_-$ = $\rho_g$ and $\rho_+$ =
$\rho_s$.  We assume that the critical value of $\rho_s$ lies on the
spinodal curve $\rho_{spin} = 1 + 2 \sqrt{\theta_c-\theta}$.  This allows us to
obtain $\kappa_{m}$, and substituting this expression into
(\ref{width}) gives a minimum interface width of
\begin{eqnarray}
w_{min} = \frac{1}{\sqrt{1+\sqrt{3}}}. 
\end{eqnarray}
As the temperature is decreased in the inset of Figure \ref{thetaw} the
critical density $\rho_{crit}$ makes a discontinuous jump to a regime in
which it lies close to the gas density, $\rho_g$.  The minimum interface
width can, in this case, be analytically obtained by expanding
densities around $\rho_g$. We define $\rho_s = \rho_g + \delta \rho$ and $\rho_+ = \rho_- + \Delta
\rho$.  Since $P-p_b$ is a positive quantity, a suitable choice for $\rho_-$
is $\rho_- = \rho_s$.  Substituting these expressions into equation
(\ref{kappamin}) gives
\begin{eqnarray}
\kappa_{m} = \max_{\delta\rho>0}\min_{\Delta\rho>0}\frac{\theta \delta \rho}{(\rho_g + \delta \rho)\Delta \rho - \Delta \rho^2/8},
\label{kappamin2}
\end{eqnarray}
Minimizing this with respect to $\Delta\rho$ leads to $\rho_s =
\Delta\rho/4$. Re-substituting this result back into equation
(\ref{kappamin2}), and maximizing with respect to $\delta\rho$, we finally
obtain $\rho_{crit} = 2 \rho_g$.  Using this we can calculate the minimum
interface width,
\begin{eqnarray}
w_{min} = \frac{1}{\sqrt{4 \rho_g \left(\theta_c - \theta \right)}},
\end{eqnarray}
as shown by the triangles in Figure \ref{thetaw}. This closely follows
the numerical result at low temperatures. This means that we need wide
interfaces for deep quenches because of the unfortunate cancellation
of the discrete derivative and Laplace operator for low densities in
the denominator of (\ref{kappamin}).

\begin{figure}
\includegraphics[width=0.8\columnwidth]{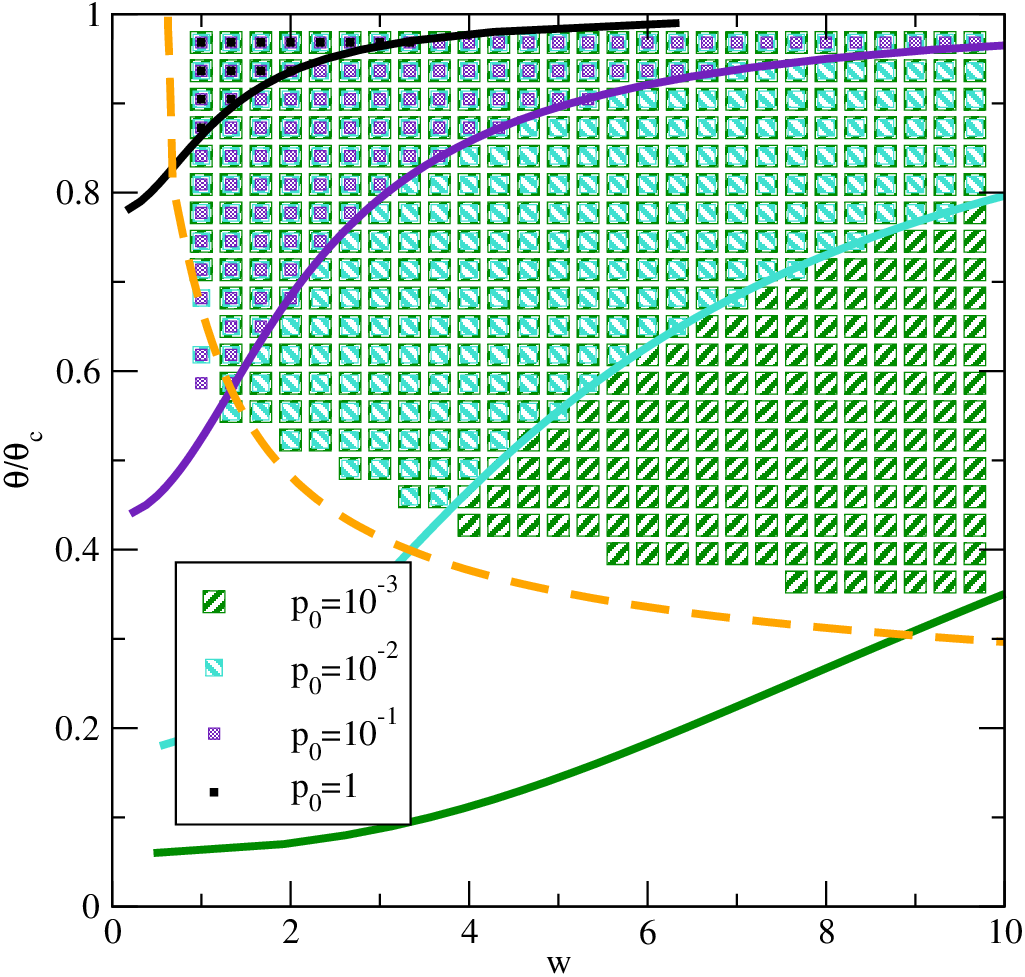}
\caption{(Color online) Existence of accurate solutions 
  for different values of $p_0$ and
  $w$. Symbols indicate parameter combinations that lead to stable,
  accurate, and unique solutions. Solid lines are the bulk stability
  limits for the pressure method given by Eq. (\ref{stabeqn}). The
  dashed line is the line for an accurate interface representation
  given by Eq. (\ref{kappamin}).}
\label{Stab_1} 
\end{figure}

Most previous lattice Boltzmann simulations approached the simulation
of non-ideal systems by using the ideal gas equation of state
$p=\rho\theta=\rho/3$, as a starting point.  Interactions are then
included to allow the simulation of non-ideal systems. The speed of
sound $c_s=\sqrt{\partial_\rho p}$ will then recover the ideal gas
value of $1/\sqrt{3}$ in the dilute limit. For a van der Waals gas
with a critical density of 1 and a temperature of $\theta=1/3$ and an
interfacial free energy of $\int \kappa/2\;(\nabla\rho)^2$ the
pressure tensor used by previous approaches matched the ideal gas
equation of state in the dilute limit, leading to $p_0=1$.  For the
van der Waals gas the speed of sound increases rapidly for high
densities. A problem arises when the speed of sound becomes larger
then the lattice velocity $|v_i|$, because information can not be
passed on at speeds larger than the lattice velocity. When the speed
of sound is increased above 1 the simulation becomes unstable. This
problem is exacerbated by the presence of the gradient terms in the
pressure tensor.  These terms further decrease the stability, as shown
in a previous analysis of the pressure method by C. Pooley for one,
two, and three dimensional lattice Boltzmann methods
\cite{PooleyThesis}. In the notation of this letter the linear
stability condition is
\begin{equation}
c_s < \sqrt{1-4p_0\kappa \rho}
\label{stabeqn}
\end{equation}
for a homogeneous, one dimensional system with density $\rho$. This
implies a restriction for both the maximum quench depth and the
maximum intereface witdth. This is shown in Fig. \ref{Stab_1} as solid
lines for different values of $p_0$.  This suggests that, at least as
far as the stability of the bulk phase is concerned, the most stable
solutions should be found for $\kappa=0$.

For simplicity we demonstrate the constraints of liquid-gas lattice
Boltzmann simulations by the common one dimensional projection of one,
two and three dimensional models. This is the D1Q3 model. The lattice
Boltzmann equation for densites $f_i$ corresponding to velocity $v_i$
is given by
\begin{equation}
f_i(\mb{x}+\mb{v}_i,t+1)=
f_i(\mb{x},t)+\frac{1}{\tau}(f_i^0(\mb{x},t)-f_i(\mb{x},t)).
\label{LB}
\end{equation}
The $f_i^0$ are the equilibrium distribution and are given by
\begin{displaymath}
f_{-1}= -\rho u/2+\Pi/2;\;f_{0}= \rho-\Pi;\;f_{1}= \rho u/2+\Pi/2
\end{displaymath}
and $\Pi=\rho u^2+P+\nu u\partial_x\rho$ \cite{Holdych,pre} where $\nu=(\tau-0.5)/3$. To second
order the resulting equations of motion are, as usual, the continuity equation
\begin{equation}
\partial_t \rho+\partial_x(\rho u)=0
\end{equation}
and the Navier Stokes equation
\begin{equation}
\partial_t(\rho u)+\partial_x(\rho u^2)=
-\partial_x P+\partial_x(2 \nu \rho \partial_x \rho)
\label{NSeqn}
\end{equation}

\begin{figure}

\includegraphics[width=0.8\columnwidth]{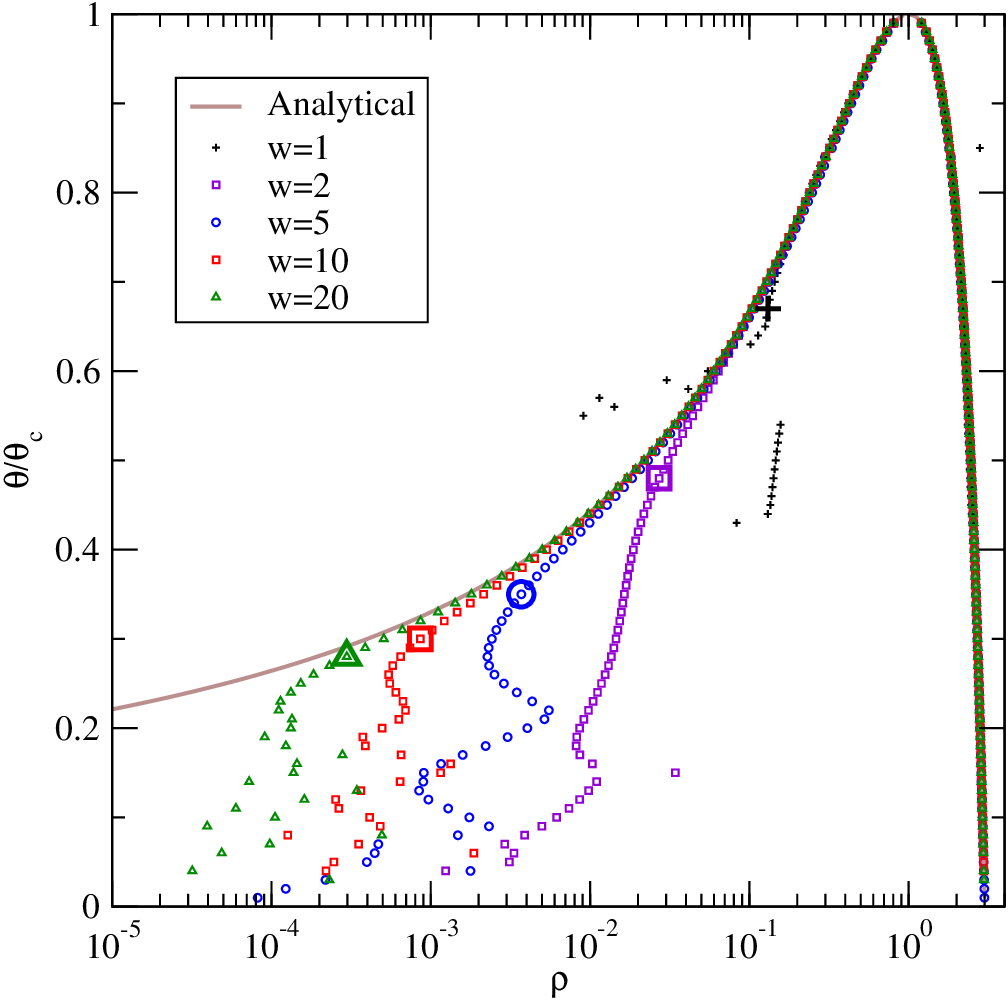}

\caption{(Color online) The van der Waals phase diagram is recovered to very good
approximation for interfaces wider than the minimum width.  The large
symbols represent the predicted points at which the simulations are
predicted to become inaccurate by Eq. (\ref{kappamin}).  The value of
$p_0$ only affects the bulk stability and values from 1 to $10^{-7}$ were
used for increasing quench depth. }
\label{FigPhase}
\end{figure}

To lower the speed of sound in the liquid phase we now reduce the value
of $p_0$ in (\ref{VdW_int_new}). This decreases the speed of sound in
the liquid by a factor $\sqrt{p_0}$. This also increases the range of
stability for $\kappa$ in (\ref{stabeqn}). We now expect that lowering the speed of
sound by a sufficient factor will reduce the speed of sound
sufficiently to simulate systems with arbitrarily low temperature
ratios $\theta/ \theta_c$. 

To test this idea we performed simulations with near equilibrium
profiles using a one dimensional three velocity $v_i=\{-1,0,1\}$ model
by defining an initial density profile that is given by two domains
with densities $\rho_l$ and $\rho_g$ respectively connected by an
equilibrium interface given by Eq. (\ref{profile}).  This profile is
not the exact analytical solution to the differential equation $\nabla
P=0$, but it is very close to it.  By initializing the simulation with
this profile we test the linear stability of the method around an
equilibrium profile to good accuracy. The shape of a stable
interfacial profile is independent of $p_0$.

In Figure \ref{FigPhase} we see that by lowering $p_0$ the method is
now able to simulate very small values of the reduced temperature $\theta/
\theta_c$ for interface width $w>1$, but that significantly
larger widths are required to recover an accurate phase diagram for
deep quenches. For values of $\theta/ \theta_c$ between 0.9 and 1 we also find
non unique solutions for small values of $\kappa$ which is discussed in
more detail in a previous paper \cite{pre}.

We now test the predictions of the accuracy of extended interface
liquid-gas simulations using a lattice Botzmann implementation first
presented in \cite{pre}.

For small values of $\kappa$ the interface becomes sharp in the continuous
limit so that the derivatives become arbitrarily large. But in the
numerical implementation the derivatives are {\em discrete}. The
discrete values are limited by the lattice spacing. In the one
dimensional case we choose $\nabla\rho(x) =0.5(\rho(x+1)-\rho(x-1))$ and
$\nabla^2\rho(x)=\rho(x+1)-2\rho(x)+\rho(x-1)$.

The methods always lead to a constant pressure, even across an
interface\cite{pre}. 
We performed a scan of the parameter space $w$ and $\theta/ \theta_c$
initializing the simulation with a near equilibrium profile for
different values of $p_0$. We accept simulations that are stable,
accurate and unique. The criterion of accuracy is defined to be
$\log_{10}(\rho_{min})-\log_{10}(\rho_g)<0.1$. As can be seen in Figure
\ref{FigPhase}, the results are not very sensitive to the exact value
of the cutoff. For values of the interface width $w<1.5$ we also test
the uniqueness of the simulation by using initial profiles with bulk
densities corresponding to the pressure at the spinodal
points\cite{pre}. Our criterion for uniqueness is that all
simulations lead to the same minimum density to within $\Delta\rho<0.01$.

Comparing (\ref{kappamin}), shown as a dashed line in Figure
\ref{Stab_1}, and the numerical results for stable, accurate and
unique solutions shows excellent agreement. The bulk stability of
Eq. (\ref{stabeqn}) gives the second limit for the acceptable
parameter range for the pressure method.  We perforemd a similar
analysis for the forcing method of \cite{pre} and obtained nearly
identical results except for a slightly larger range of bulk
stability. Note that previous lattice Boltzmann simulations use
$p_0=1$, which corresponds to the area under the black line in
Fig. \ref{Stab_1}. This is why it was assumed that lattice Boltzmann
simulations are limited to a maximum density ratio of about 10
\cite{Inamuro}.

The interface constraint (\ref{kappamin}) is remarkably successful at
predicting the acceptable simulation parameters. It predicts how thin
is too thin for an interface. It thereby detects when non-unique
solutions occur and when solutions for deep quenches fail to deliver
accurate results.

For simulation methods it is important to be aware of the acceptable
paramter ranges. Lattice Boltzmann simulations are often believed to
have the nice property of becoming unstable before they become
inaccurate \cite{cates}. This is not the case for thin interfaces in
multi-phase simulations. In this case the simulation can remain stable
and become inaccurate. This makes it necessary to find some criterion
that determines whether a set of paramters will lead to an accurate
simulation. Such a criterion was presented for the interface width
(controlled by $\kappa$) in this communication.

The second contribution presented in this paper appears trivial at
first: in consists of a simple pre-factor for the pressure. It is new,
because it breaks with the idea that the standard lattice Boltzmann
method for ideal gases should always be recovered in the dilute
limit. This prefactor has profound implications on the stability of
the bulk phases, which can be seen using an important result about the
bulk stability from \cite{PooleyThesis}.

Combining these two components we were able to show that lattice
Boltzmann is indeed able to simulate very deep quenches for liquid-gas
cases. This analysis was general and will be applied to other
equation of states as well as other discretization of the interfacial
terms. This may yet yield significant further advances for the
development of multi-phase lattice Boltzmann methods.

\end{document}